\documentclass[sigplan,nonacm]{acmart}
\usepackage[utf8]{inputenc}
\usepackage{natbib}
\usepackage[noframes,novert]{ffcode}
\usepackage{afterpage}
\usepackage[off]{to-be-determined}
\usepackage{float}
\usepackage{href-ul}
\usepackage{paralist}
\usepackage{caption}
\usepackage{bibcop}
\usepackage{doi}
\usepackage{soul}
\usepackage[capitalize]{cleveref}
\usepackage{booktabs}
\usepackage{tabularx}
\usepackage{graphicx}
\usepackage{pbox}
\usepackage{xcolor}
\usepackage{float}
\usepackage{multicol}

\title{Evaluating the Dependency Between Cyclomatic Complexity and Response For Class}

\author{Maxim Stavtsev}
\email{mastavtsev@edu.hse.ru}
\orcid{0009-0008-8171-8167}
\affiliation{\institution{Higher School of Economics}\city{Moscow}\country{Russia}}

\author{Yegor Bugayenko}
\orcid{0000-0001-6370-0678}
\email{yegor256@gmail.com}
\affiliation{\institution{Huawei}\city{Moscow}\country{Russia}}

\begin{document}

\begin{abstract}
In object-oriented programming, it is reasonable to hypothesize that smaller classes with fewer methods are less complex. Should this hypothesis hold true, it would be advisable for programmers to design classes with fewer methods, as complexity significantly contributes to poor maintainability. To test this assumption, we analyzed 862,517 Java classes from 1,000 open GitHub repositories. Our findings indicate a strong Pearson correlation of 0.79 between the cumulative McCabe's Cyclomatic Complexity (CC) of all class methods and the number of methods, a metric known as Response for Class (RFC).
\end{abstract}

\maketitle

\section{Introduction}

\tbd{Problem 1: The recommendation provided by the authors is partially based on the results regarding Pearson's coefficient which I see critical. Using Pearson in hypothesis testing requires both variables to be close to normally distributed, moreover
homoscedasticity should be given. Figure 1 and 2 show that neither is the case. So the results according to Spearman remain showing a moderate correlation which, in my opinion, is not enough to support such a recommendation.}

\tbd{Problem 2: From what the authors presented, it does not seem that the conducted study targets an interesting research problem. To be interesting, the answer to a research hypothesis should not be apparent before conducting the study. However, in this case, it seems quite apparent that increasing RFC increases CC, i.e., it seems that such hypothesis cannot be falsified.}

\tbd{Problem 3: Using two different correlation coefficients on the same data is inappropriate (as each has different assumptions about the data distribution, which cannot be true simultaneously!). Also one needs to be careful how to report and discuss p-values as well as statistic values.}

Software complexity has long been a focal point in software engineering, given its substantial influence on the maintainability, readability, and overall quality of code~\citep{li1993object, yamashita2016thresholds}. Among the metrics developed to assess software complexity, Cyclomatic Complexity (CC) introduced by \citet{mccabe1976complexity} stands out as a widely accepted indicator. It quantifies complexity by measuring the number of linearly-independent paths through the source code.

The question arises as to what specifically leads to complexity, or more precisely, which metrics correlate with CC? Previous studies have identified a correlation between CC and various factors such as the size of a class in lines of~\citep{meine2007correlations, graylin2009cyclomatic}, its depth of indentation~\citep{hindle2008reading}, and its inheritance depth~\citep{seront2005relationship}. However, there has been limited research on how the Response for Class (RFC)---the number of methods in a class~\citep{chidamber1994metrics}---correlates with CC, while numerous books on programming suggest avoiding classes with too many methods~\citep{mcconnell2004code, martin2008clean}. Finding a correlation here might give a more grounded suggestion to programmers that writing classes with fewer methods could enhance maintainability.

In this study, we analyzed 862,517 Java classes from 1,000 open GitHub repositories, calculating their CC and RFC metrics. We applied both Pearson and Spearman correlation methods to explore the relationship between these metrics. Our findings reveal a strong Pearson correlation of 0.79 and a moderate Spearman correlation of 0.59 between CC and RFC. In both instances, \(p\)-values were less than 0.001, indicating statistical significance.

This article is structured as follows: 
\cref{sec:related} presents a review of works related to the research, 
\cref{sec:method} outlines practical steps taken during the study, 
\cref{sec:experiment} explains the details of our experiment, 
\cref{sec:results} describes obtained results, 
\cref{sec:discussion} provides interpretations of our findings and explores limitations, and 
\cref{sec:conclusion} offers a summary of the paper.

\section{Related Work}\label{sec:related}

We are not the first, who analyze the correlation between CC and RFC. Recently, \citet{mamun2017correlations} showed that there is a strong correlation between RFC and CC, which lead them to conclusion that one metric can be a proxy for the other. Moreover, since CC is harder to measure, it could be eliminated. However, the dataset employed by the authors might be considered limited and potentially biased, as it contained only 20 GitHub repositories. Additionally, the authors did not categorized the Java methods that they analyzed, such as object methods and static methods.

There were studies analyzing correlation between CC and some other metrics. For example, \citet{meine2007correlations} and \citet{graylin2009cyclomatic} demonstrated the presence of a strong linear correlation between CC and Lines of Code (LoC), while \citet{landman2016empirical} demonstrated that for CC and LoC, the correlation is not strong enough to conclude that CC is redundant with LoC; \citet{muslija2018correlation} showed that a correlation between the effort needed to test a program and its complexity is moderate; \citet{seront2005relationship} observed no significant correlation between the depth of inheritance of a class and its weighted method complexity; \citet{abd2018} showed that the probability of the occurrence or emergence of new errors positively correlates with the CC; \citet{shin2008complexity} found a weak correlation between code vulnerability and its complexity; \citet{hindle2008reading} discovered a correlation between CC and the depth of indentation; \citet{mamun2017correlations} identified a correlation between complexity and the quality of source code documentation.

To the best of our knowledge, no study has closely investigated the relationship between CC and RFC employing both statistical analyses and graphical representations.

\section{Method}\label{sec:method}

The goal of this study was to explore whether Java classes with smaller number of methods are less complex. This lead us to following research question: Is there a correlation between CC and RFC?

First, we took the CaM dataset\footnote{\url{https://github.com/yegor256/cam}} as one of the primary instruments for our research~\citep{cam2024}. The idea behind CaM project is to create a standardized archive that had already done much of the preliminary work, including mining, filtering, and collecting metrics for Java code from open-source projects. To generate the data in CaM, their authors wrote Bash and Python scripts that performed the following functions: 
\begin{enumerate} 
  \item They mined Java repositories from GitHub, specifically targeting those tagged with the "\ff{java}" label, which had a star count between 1K and 10K, and had a size at least as large as 200Kb.
  \item  They filtered out non-essential files, removing those without the "\ff{.java}" extension, files with syntax errors, certain metadata files like "\ff{package-info.java}", files with excessively long lines, and unit tests.
  \item After the filtration process, their scripts calculated numerous metrics for each Java class. These metrics encompassed various aspects of code quality and structure, such as LoC, Non Commenting Source Statements (NCSS), CC, Cognitive Complexity, and the number of different class components like attributes, constructors, and methods. 
\end{enumerate}
We used the ``2023-10-22'' version of the CaM dataset (2.19Gb). It contained 862,517 Java classes from 1,000 GitHub open repositories. We believe that the method is ethical, as it utilizes data from publicly available sources, thereby avoiding any infringement of copyright.

Yet another benefit of the CaM repository is that it aggregates 33 pre-calculated metrics, including CC, RFC, SRFC (only static methods are counted), and NSRFC (only non-static methods are counted). By utilizing the CaM repository, we leveraged a ready-made archive that not only ensured the repeatability of our research results but also saved considerable time in data pre-processing. It is crucial to describe the method utilized for calculating CC by the creators of the CaM repository. They employed a Python library designed to analyze Java source code, known as javalang\footnote{\url{https://github.com/c2nes/javalang}}. The computation of CC is predicated on constructing an Abstract Syntax Tree (AST) and subsequently investigating each node within the tree. The complexity count is incremented by one whenever an AST node corresponds to a binary operation involving logical conjunctions ('\&\&') or disjunctions ('||'), any control flow statement (including 'if', 'for', 'while', 'switch'), or a 'try' statement.

Then, we conducted a comparative analysis of the CC metric across classes with varying numbers of methods. We constructed graphical representations, such as histograms, box plots, to depict the distribution of CC in classes with similar RFC, and scatter plots to illustrate the overall trend of CC as the RFC changes. Furthermore, we leveraged the capabilities of Scikit-learn\footnote{\url{https://scikit-learn.org/}}, one of the most popular Python libraries for machine learning with over 56,000 stars on GitHub, to model the associations between CC and RFC in a Java class. Our analysis explored both the classical Pearson correlation followed by linear regression and Spearman's correlation, which is based on ordered statistics, to represent the underlying relationships.

\section{Experimental Setup}\label{sec:experiment}

Before delving into the results, we will initially explore the selected visualizations and statistical methods.

We illustrated the distributions of CC and RFC through histograms and descriptive statistics (median, mean, min, max). Additionally, we created a box plot for each RFC category to depict the CC distribution within each category.

Simple scatter plot with RFC on the horizontal axis and CC on the vertical axis were not suitable for our data. Due to long tail distribution of both metrics, the data was concentrated in the lower left part of the plot and many of the dots are placed on top of each other. We utilized the approach of hexagonal scatter plots~\citep{carr1987scatterplot} to get rid of overplotting. The latter method segmented the plot area's two-dimensional plane into a grid of 15x15 hexagons. This method tallies the number of data points within each hexagon, coloring them based on a logarithmic 255-step violet scale gradient to represent density. Hexagonal plots, by this design, offered a clearer visual representation than traditional scatter plots, significantly reducing confusion. 

In our analysis, we employed both Pearson and Spearman correlation coefficients to assess the relationship between CC and RFC. The Pearson correlation coefficient measured the linear relationship between these two metrics, providing insight into how closely changes in one metric are associated with changes in the other. On the other hand, the Spearman correlation coefficient did not assume a normal distribution and instead evaluated the monotonic relationship between the two variables.

Finding a strong Pearson's correlation coefficient allowed us to develop a reliable predictive linear model between the two variables, potentially making one of the metrics unnecessary.  However, for other types of correlation measures, such as Spearman's method, the connection between regression and correlation was not inherently straightforward.

\section{Results}\label{sec:results}

\begin{figure*}
  \centering
  \includegraphics[scale=0.4]{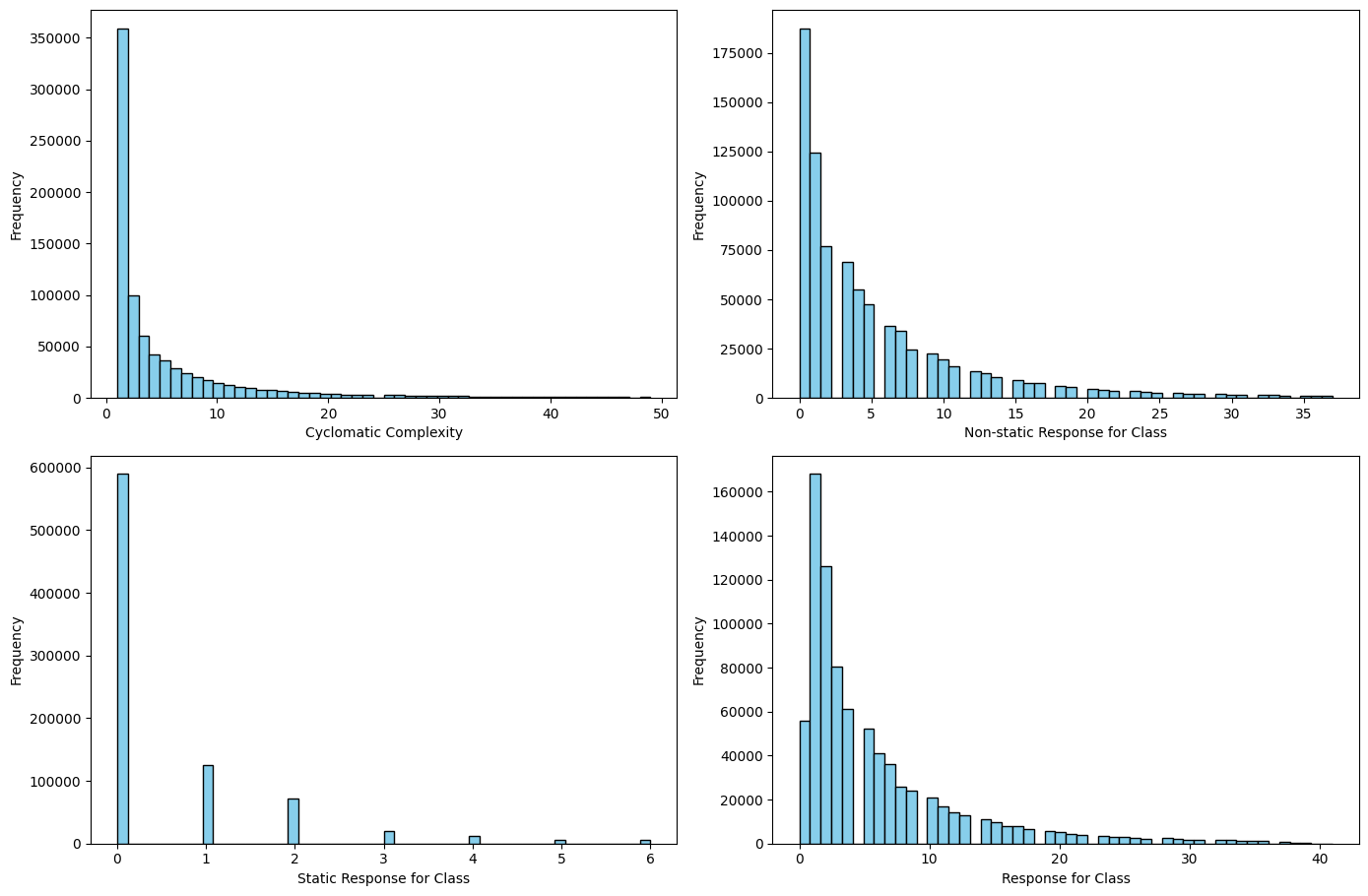}
  \caption{Evidence of right-skewness in metrics. The histograms display the frequency distribution of metrics. Each metric's value is plotted on the x-axis, while the corresponding number of classes is presented on the y-axis. This visualization clearly depicts a right-skewed tendency, signifying a predominance of classes with lower value of metrics within the data. }
  \Description{
    A set of four histograms displaying frequency distributions for various software complexity metrics. The top-left histogram shows the distribution of Cyclomatic Complexity, with the majority of the values concentrated at the lower end. The top-right histogram presents the distribution of Non-static Response for Class, with a similar skew towards lower values. The bottom-left histogram details the distribution of Static Response for Class, which is even more sharply skewed to the left. The bottom-right histogram illustrates the distribution of Response for Class, combining both static and non-static methods, and shows a consistent pattern of skewness towards lower values.
    }
  \label{fig:distributions}
\end{figure*}

\Cref{fig:distributions} graphically presents the distributions of CC, RFC, NSRFC, and SRFC among all Java classes. \Cref{tab:method_stats} shows the statistics of the distribution. Based on CC, there are many examples with values ranging from 0 to 4, after which the number of examples declined rapidly. A similar rapid decline was observed for other metrics as well. Thus, we concluded that the data exhibits right-skewness. 

\begin{table}
\caption{Summary statistics for Cyclomatic Complexity and Response for Class Metrics. The metrics reveal a general trend: a significant number of observations cluster at the lower end of the scale, yet there are instances of high complexity or number of methods, as demonstrated by the maximums. The mean values prove a right-skewed distribution. The quartile data further illustrates the concentration of lower values with a right skew.}
\label{tab:method_stats}
\begin{tabularx}{\linewidth}{X>{\ttfamily}r>{\ttfamily}r>{\ttfamily}r>{\ttfamily}r>{\ttfamily}r>{\ttfamily}r}
\toprule
Metric & {\rmfamily Min} & {\rmfamily 25\%} & {\rmfamily Median} & {\rmfamily Mean} & {\rmfamily 75\%} & {\rmfamily Max} \\
\midrule
CC     & 1.0 & 1.0  & 2.0    & 5.46 & 6.0  & 49.0 \\
NSRFC  & 0.0 & 1.0  & 3.0    & 5.24 & 7.0  & 37.0 \\
SRFC   & 0.0 & 0.0  & 0.0    & 0.53 & 1.0  & 6.0  \\
RFC   & 0.0 & 1.0  & 3.0    & 5.76 & 8.0  & 41.0 \\
\bottomrule
\end{tabularx}
\end{table}

\begin{figure*}
\includegraphics[scale=0.4]{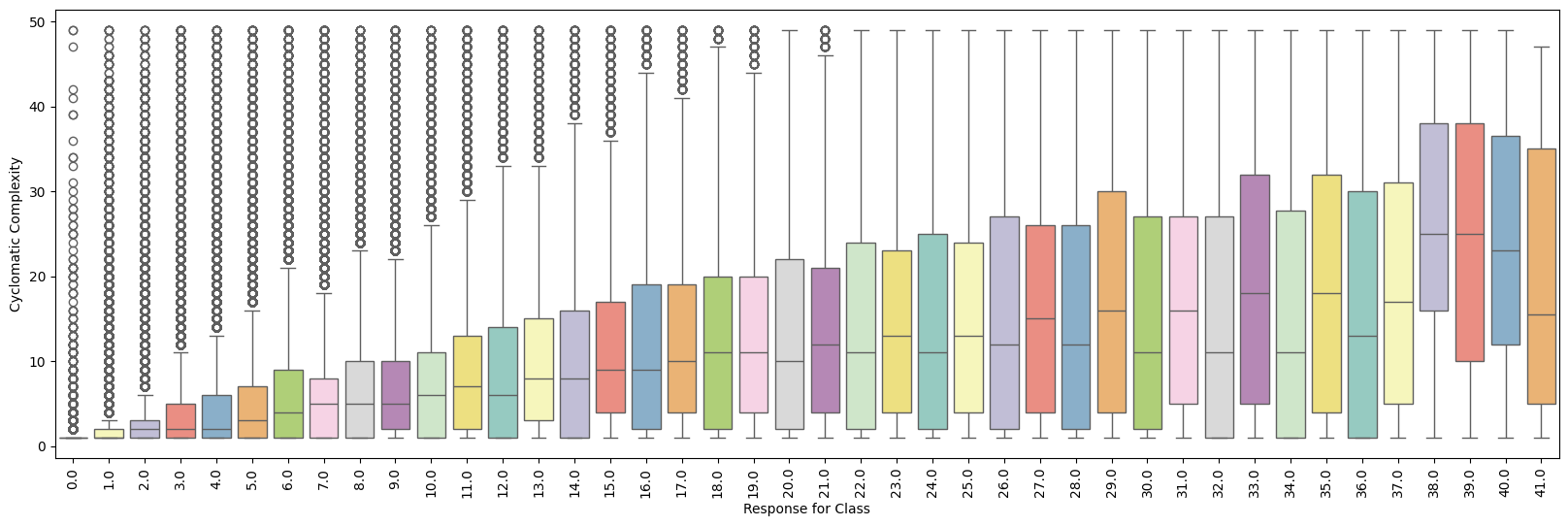}
\caption{Augmented variability in Cyclomatic Complexity associated with increased Response for Class. Box plots depicts the range and median of CC for classes at increasing levels of RFC. Outliers are denoted as individual points. The analysis shows an upward trend in CC variability as RFC grow, underlining the increasing complexity in larger classes.}
\Description{
    A horizontal box plot visualization represents the relationship between Response for Class and Cyclomatic Complexity. The plot consists of a series of box plots, each corresponding to a Response for Class value from 0 to 40, showing the distribution of Cyclomatic Complexity for each class. The plot shows an increase in variance of Cyclomatic Complexity as the Response for Class increases.
    }
\label{fig:boxplots}
\end{figure*}

\Cref{fig:boxplots} illustrates the increasing variance of the CC metric as the RFC increases. Moreover, the median is increasing, but so is the inter-quartile range.

\begin{figure}
\includegraphics[scale=0.4]{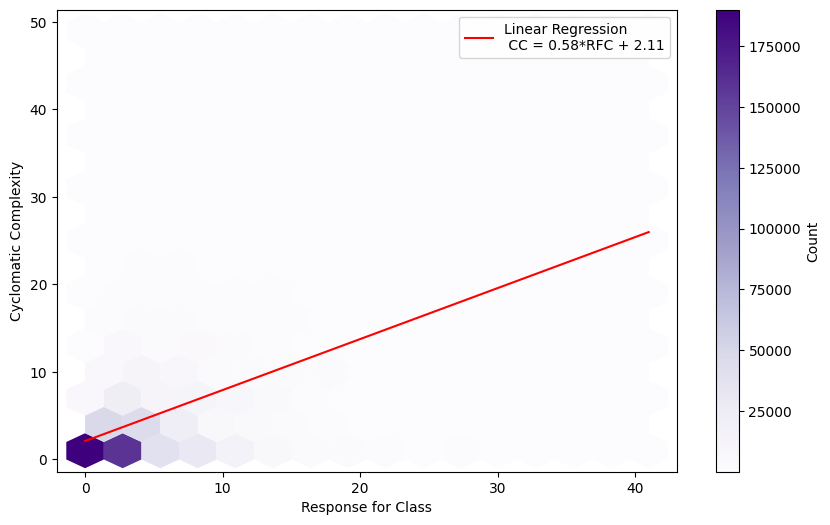}
\caption{High data density in lower ranges of Cyclomatic Complexity and Response for Class metrics. This plot highlights the commonality of simpler classes within the dataset.}
\Description{
    A hexagonal plot with a linear regression line, depicting the relationship between Response for Class on the x-axis and Cyclomatic Complexity on the y-axis. The hexagonal bins display the count of occurrences with darker colors representing higher frequencies. A prominent red line representing a linear regression model is shown on the plot, with the equation 'CC = 0.58*RFC + 2.11' indicating a positive linear relationship.
    }
\label{fig:hexa_plot}
\end{figure}

\Cref{table:pearson} and \cref{table:spearman} present the Pearson and Spearman correlation coefficients between CC and various types of method counts, along with their respective \(p\)-values. \Cref{table:pearson} shows the Pearson correlation, which assesses the linear relationship between CC and each method count type. It reveals a strong positive correlation of 0.79 for both NSRFC and RFC, and a slightly lower, but still substantial, correlation of 0.63 for SRFC, with all \(p\)-values at 0.0, indicating statistical significance. \Cref{table:spearman} presents the Spearman correlation coefficients, reflecting the monotonic relationships. Here, the correlations are notably lower, with NSRFC at 0.49, SRFC at 0.24, and RFC at 0.59, again with \(p\)-values at 0.0, suggesting significance. 

The linear Pearson correlation coefficient surpasses the Spearman rank correlation coefficient when examining the data. This observation may be explained by the pronounced impact of outlier observations in the distribution tails of the metrics, particularly when compared to their ranked equivalents. This analysis suggests that there are key instances within software source code indicating that classes characterized by high values of RFC, NSRFC, and SRFC tend to exhibit increased complexity. Conversely, the Spearman correlation, noted for its robustness to outliers, reveals a moderate correlation between RFC and CC, with a coefficient of 0.59, suggesting that  a significant correlation persists even when outliers are less emphasized.

\Cref{fig:hexa_plot} illustrates the relationship between CC and RFC. The hexagonal plot demonstrates the substantial data density in areas of lower RFC and CC, confirming the skew towards lower complexity observed in the distribution histograms. Each hexagon represents a grouping of data points, with the color intensity reflecting the count of occurrences; darker hexagons signify a higher concentration of data points. The plot also features a linear regression line, indicating a positive relationship with the equation \(\text{CC} = 0.58 \times \text{RFC} + 2.11\), suggesting that as the number of methods increases, so does the CC. This equation is obtained from the coefficients of a linear regression model fitted to our data via the Scikit-learn module.

\begin{table}
\caption{Pearson correlation between Cyclomatic Complexity and RFC metrics within software systems. The coefficients suggest strong positive correlations between CC and both NSRFC and RFC (0.79), and a moderately strong correlation with SRFC (0.63), all with \(p\)-values less than 0.001.}
\label{table:pearson}
\begin{tabularx}{\linewidth}{X>{\ttfamily}r>{\ttfamily}r}
\toprule
Metric                     & {\rmfamily Pearson Coefficient} & {\rmfamily \(p\)-value} \\
\midrule
NSRFC         & 0.79                & <0.001  \\
SRFC             & 0.63                & <0.001  \\
RFC              & 0.79                & <0.001  \\
\bottomrule
\end{tabularx}
\end{table}

\begin{table}
\caption{Spearman correlation between Cyclomatic Complexity and RFC metrics within software systems. The results indicate a moderate correlation between CC and RFC and CC and NSRFC, with coefficients of 0.59 and 0.49, respectively. In contrast, a weaker correlation exists between CC and static SRFC, evidenced by a coefficient of 0.24.}
\label{table:spearman}
\begin{tabularx}{\linewidth}{X>{\ttfamily}r>{\ttfamily}r}
\toprule
Metric                     & {\rmfamily Spearman Coefficient} & {\rmfamily \(p\)-value} \\
\midrule
NSRFC         & 0.49                 & <0.001  \\
SRFC             & 0.24                 & <0.001  \\
RFC              & 0.59                 & <0.001  \\
\bottomrule
\end{tabularx}
\end{table}

\section{Discussion}\label{sec:discussion}

Based on our dataset and the calculations we conducted, we obtained results that contrast with those presented by \citet{mamun2017correlations}. We observed only a moderate Spearman correlation between RFC and CC, quantified by a coefficient of 0.59, whereas \citet{mamun2017correlations} reported a much higher coefficient of 0.989, with a \(p\)-value of 0.0. Furthermore, our study extended the analysis to static and non-static methods separately, which was not conducted in the \citet{mamun2017correlations} study. This separate examination yielded even lower correlations with CC, recording coefficients of 0.24 and 0.49, respectively.

CC appears to depend linearly on all three types of method counts according to the Pearson correlation, and it does not exhibit monotonic relations according to the Spearman correlation. On the one hand, this suggests that the CC metric may be redundant within the software quality measurement process if any RFC metric is already used, since they linearly depend and reflect similar patterns in the code. On the other hand, if the goal is to write less complex and thus more maintainable code, attention should be given to the number of methods in a class; as this number increases, so does the complexity of the code.

There are certain limitations in our work. Firstly, our analysis does not include proprietary repositories, potentially omitting data that could influence the generalizability of our findings. Secondly, our approach does not involve manual inspections of the classes or the filtering out of corner cases, factors that could potentially compromise the precision of our correlation assessments. Future studies should aim to address these limitations, increasing the size and diversity of the dataset, and conducting the manual inspections of the data samples.

\section{Conclusion}\label{sec:conclusion}

This study identifies a significant linear Pearson correlation of 0.79 between Cyclomatic Complexity (CC) and Response for Class (RFC) across 862,517 Java classes from 1,000 open GitHub repositories. These results challenge the practice of using CC as an independent complexity metric in Java classes without considering RFC. Consequently, we conclude that it is reasonable to advise programmers to design classes with fewer methods, as this approach can contribute to enhanced code maintainability.

\bibliographystyle{ACM-Reference-Format}
\bibliography{main}

\end{document}